\documentstyle[aps,preprint,eqsecnum,tighten]{revtex}
\begin{document}
\preprint{\today}
\title{Scaling behavior for finite
${\cal O}(n)$ systems with long-range interaction}
\author{
Hassan Chamati\footnote{Electronic address: chamati@issp.bas.bg}
and Nicholay S. Tonchev\footnote{Electronic address: tonchev@issp.bas.bg}}
\address{Institute of Solid State Physics, 72 Tzarigradsko Chauss\'ee,
1784 Sofia, Bulgaria}
\maketitle
\draft
\begin{abstract}
A detailed investigation of the scaling properties of the fully finite
${\cal O}(n)$ systems, under periodic boundary conditions, with
long-range interaction, decaying  algebraically with the interparticle
distance $r$ like $r^{-d-\sigma}$, below their upper critical dimension
is presented. The computation of the scaling functions is done to one
loop order in the non-zero modes. The results are obtained in an
expansion of powers of $\sqrt\varepsilon$, where $\varepsilon=2\sigma-d$
up to ${\cal O}(\varepsilon^{3/2})$. The thermodynamic functions are
found to depend upon the scaling variable 
$z=RU^{-1/2}L^{2-\eta-\varepsilon/2}$, where $R$ and $U$ are the
coupling constants of the constructed effective theory, and $L$ is the
linear size of the system. Some simple universal results are obtained.
\end{abstract}

\pacs{PACS numbers: 05.70.Jk, 64.60.Ak, 64.60.Fr}

\section{introduction}\label{introduction}
The theory of continuous phase transitions is based on the hypotheses
that at temperatures close to the critical $T_c$, there is only one
dominating length scale related with the critical behavior of the
system. Because of the divergent nature of the correlation length as
the critical point is approached, the microscopic details of the
system becomes irrelevant for the critical exponents describing the
singular dependence of the thermodynamic functions. This intuitive
picture is based on the grounds of the renormalization group treatment
of second order phase transitions.

Scaling is a central idea in critical phenomena near a continuous
phase transition and in the field theory when we are interested in the
continuum limit~\cite{zinnjustin96}. In both cases we are interested
in the singular behavior emerging from the overwhelming large number
of degrees of freedom, corresponding to the original cutoff scale,
which need to be integrated out leaving behind long-wave length which
vary smoothly. Their behavior is controlled by a dynamically generated
length scale: the correlation length $\xi_b$. Such a fundamental idea is
difficult to test theoretically because it requires a study of a huge
number of interacting degrees of freedom. Experimentally, however, one
hopes to be able to study scaling in finite systems near a second
order phase transition. Namely the system is confined to a finite
geometry and the finite-size scaling theory is expected to describe
the behavior of the system near the bulk critical temperature (for a
review on the finite-size scaling theory see
Ref.~\cite{barber83,privman90}).

The ${\cal O}(n)$-symmetric vector models are extensively used to
explore the finite-size scaling theory, using different methods and
techniques both analytically and numerically. The most thoroughly
investigated case is the particular one corresponding to the limit
$n=\infty$ (this limit includes also the mean spherical
model)~\cite{privman90}. In this limit, these models are exactly soluble
for arbitrary dimensions and in a general geometry. These investigations
were devoted exclusively to systems with short (including nearest
neighbors) as well as long-range forces decaying with the interparticle
distance in a power law. For finite $n$ the most frequently used
analytical method is that of renormalization
group~\cite{zinnjustin96,dohm93}. However this is limited to the case of
short-range interaction. The crossover from long to short-range forces
was discussed in Ref.~\cite{korucheva91}, where it has been found that
renormalized values of the temperature and the coupling constant are
continuous functions of the parameter controlling the range of the
interaction, when this approaches the value $2$ characterizing the
short-range force potential. The case of pure long-range interaction was
investigated very recently in Ref.~\cite{luijten99} (a comment on the
method and the results obtained there is presented in
Section~\ref{discussion}). In the mean time, a special attention was
devoted to the investigation of finite size-scaling for the
mean-spherical model with long-range interaction (for a review see Ref.
\cite{brankov92} and references therein).

In recent years there has been an increasing interest in the numerical
investigation of the critical properties of systems with long-range
interaction decaying at large distances $r$ by a power-law as
$r^{-d-\sigma}$, where $d$ is the space dimensionality and $\sigma$ is
the parameter controlling the range of the interaction. The mostly used
technique for this achievements is the Monte Carlo method. This method
was used to investigate the critical properties of Heisenberg
ferromagnetic systems~\cite{romano96} as well as Ising
models~\cite{luijten97,bayong99}. Nevertheless all the analysis there
was concentrated on systems with classical critical behavior in the
sense that the critical exponents are given by Landau theory.

In this paper we present a detailed investigation of the finite-size
scaling properties of the field theoretic ${\cal O}(n)$ vector $\varphi^4$
model with long-range interaction. We will also check the influence of
the interaction range on the critical behavior. These interactions enter
the exact expressions for the free energy only through their Fourier
transform, which leading asymptotic is $U(q)\sim q^{\sigma^*}$, where
$\sigma^*=\min(\sigma,2)$ \cite{fisher72}. As it was shown for bulk
systems by renormalization group arguments $\sigma\ge 2$ corresponds
to the case of finite (short) range interactions, i.e. the universality
class then does not depend on $\sigma$ \cite{fisher72,yamazaki77}.
Values satisfying $0<\sigma<2$ correspond to long-range interactions
and the critical behavior depends on $\sigma$. With the renormalization
group treatment it has been found that the critical behavior depend on
the small parameter $\varepsilon=2\sigma-d$, where $2\sigma$
corresponds to the upper critical dimension \cite{fisher72}. According to
the above reasoning one usually considers the case $\sigma>2$ as
uninteresting for critical effects, even for the finite-size treatments
\cite{fisher86}. So, here we will consider only the case
$0<\sigma\leq2$.

Here, we will provide a systematic and controlled approach to the
quantitative computation of the thermodynamic momenta, usually used
in numerical analysis. These momenta are related to the Binder's
cumulant and to various thermodynamic functions like the susceptibility.
We will concentrate on the scaling properties of the coupling constants
defining the system in the vicinity of the critical point. Our method is
quite general and should apply to a large extend to the investigation of
finite-size scaling in systems with long-range interaction in the vicinity
of the critical point.

%Recently, in Ref.~\cite{luijten99} the case of ${\cal O}(n)$ systems
%with long-range interaction was investigated using different
%mathematical technique. We  will comment on its results in
%section~\ref{discussion}.

The plan of the paper is as follows. In Section~\ref{secphi} we review,
briefly, the $\varphi^4$-model with long-range interaction and discuss
its bulk critical behavior. Section~\ref{secfss} is devoted to the
explanation of the methods used here to achieve our analysis. We end the
section with the computation of some thermodynamic quantities of
interest. In Section~\ref{discussion} we discuss our results briefly. In
the remainder of the paper we present some details of the calculations
of some formula used throughout the paper.

\section{Finite-size scaling for systems with long-range interactions}
\label{secphi}
In the vicinity of its critical point the Heisenberg model, with
long-range interaction decaying as power-law, is equivalent to the
$d$-dimensional ${\cal O}(n)$-symmetric model~\cite{suzuki73}
\begin{equation}\label{model}
\beta{\cal H}\left\{\varphi\right\}=\frac12\int_V d^dx\left[
\left(\nabla^{\sigma/2}\varphi\right)^2+r_0\varphi^2
+\frac12u_0\varphi^4\right],
\end{equation} where $\varphi$ is a short hand notation for the space
dependent $n$-component field $\varphi(x)$, $r_0=r_{0c}+t_0$
($t_0\propto T-T_c$) and $u_0$ are model constants. $V$ is the
volume of the system. In equation~(\ref{model}), we assumed
$\hbar=k_B=1$ and the size scale is measured in units in which the
velocity of excitations $c=1$. We note that the first term in the model
denotes ${\bbox k}^\sigma|\varphi({\bbox k})|^2$ in the momentum
representation where the parameter $0<\sigma\leq2$ takes into
account short-range as well as long-range interactions. $\beta$ is the
inverse temperature. The nature of the spectrum suggests that the
critical exponent $\eta=2-\sigma$ \cite{fisher72,yamazaki77}. Here we
will consider periodic boundary conditions. This means
\begin{equation}
\varphi(x)=\frac1{\sqrt{V}}\sum_{\bbox k}\varphi({\bbox k})
\exp\left(i{\bbox k}\cdot x\right),
\end{equation}
where ${\bbox k}$ is a discrete vector with components $k_i=2\pi
n_i/L$ $\left(n_i=0,\pm1,\pm2,\cdots,\ i=1,\cdots,d\right)$ and a cutoff
$\Lambda\sim a^{-1}$ ($a$ is the lattice spacing). In this paper, we
are interested in the continuum limit i.e. $a\to0$. As long as the
system is finite we have to take into account the following
assumptions $L/a\to\infty$, $\xi_b\to\infty$ while $\xi_b/L$ is finite.

Fisher {\it et al.}~\cite{fisher72} and Yamazaki {\it et
al.}~\cite{yamazaki77} have shown that for the model under
consideration the Landau theory holds for $d>2\sigma$. In the
opposite case i.e. $d<2\sigma$ an expansion in powers of
$\varepsilon=2\sigma-d=4-d-2\eta$ takes place, where $2\sigma$
plays the role of the upper critical dimension. We will present the
renormalized parameters which characterize the bulk critical behavior
and appear in the scaling functions. Since the computations are
standard~\cite{zinnjustin96}, we will be quite brief.

The application of the renormalized theory, above the critical
temperature, to the model Hamiltonian requires a scaling field amplitude
$Z$, a coupling constant renormalization $Z_g$ and a renormalization of
the $\varphi^2$ insertions in the critical theory $Z_t$. In term of
these, we define as usual
\begin{equation}\label{definition}
t=Z Z_t^{-1}(r_0-r_{0c}) \ \ \ \ \text{and } \ \ \ \
g=\ell^{-\varepsilon}Z^2Z_g^{-1}u_0.
\end{equation}
In the remainder we will work in units where the reference length $\ell$
is set to unity. To one loop order the renormalization constants in the
minimal subtraction scheme are given by~\cite{yamazaki77}
\begin{mathletters}\label{renormalization}
\begin{equation}\label{Z}
Z=1+{\cal O}(\hat g^2)
\end{equation}
\begin{equation}\label{Zt}
Z_t=1+\frac{n+2}\varepsilon\hat g+{\cal O}(\hat g^2)
\end{equation}
\begin{equation}\label{Zg}
Z_g=1+\frac{n+8}\varepsilon\hat g+{\cal O}(\hat g^2)
\end{equation}
\end{mathletters}
In Eqs.(\ref{renormalization}),
\begin{equation}\label{hatg}
\hat g=g\frac{2}{(4\pi)^{d/2}\Gamma(d/2)}=\frac{2g}{(4\pi)^\sigma
\Gamma(\sigma)}\left(1+\frac\varepsilon2\left[\ln(4\pi)+
\psi(\sigma)\right]+{\cal O}(\varepsilon^2)\right),
\end{equation}
where $\psi(x)$ is the digamma function.

The fixed point of the $\beta$ function is at $\hat g=\hat g^*$ with
\begin{equation}\label{fixed}
\hat g^*=\frac\varepsilon{n+8}+{\cal O}(\varepsilon^2).
\end{equation}

Before starting to investigate the finite-size scaling in the field
theoretical model under consideration, we shall recall briefly the
corresponding renormalization group formalism. In the continuum limit,
the lattice spacing completely disappears. The integration over wave
vectors of the fluctuations are evaluated without cutoff and are
convergent. When some dimensions of the system are finite the
integrals over the corresponding momenta are transformed into sums.
Since the lattice spacing is taken to be zero, the limits of the sums
still tend to infinity.

From general renormalization group considerations an observable $X$,
the susceptibility for example, will scale like \cite{brezin82}:
\begin{equation}\label{chi}
X[t,g,\ell,L]=\zeta(\rho)X\left[t(\rho),g(\rho),\ell\rho,L\right],
\end{equation}
where $t$ is the reduced temperature, $g$ a dimensionless coupling
constant and $L$ the finite-size scale. The length scale $\ell$ is
introduced in order to control the renormalization procedure.

It is known that in the bulk limit, when $g(\rho)$ approaches its stable
fixed point $g^*$ then we have
\begin{equation}\label{rho}
t(\rho)\approx t\rho^{1/\nu} \ \ \ \ {\rm and} \ \ \ \
\zeta(\rho)\approx
\rho^{\gamma_x/\nu},
\end{equation}
where $\gamma_x$ and $\nu$ are the bulk critical exponents
measuring the divergence of the observable $X$ and the correlation
length, respectively, in the vicinity of the critical point and $\rho$ is a
scaling parameter. Using dimensional analysis together with
equation~(\ref{chi}) one gets
\begin{equation}\label{fss}
X[t,g,\ell,L]=\zeta(\rho) X\left[t(\rho)(\rho
\ell)^2, g(\rho),1,L/\ell\rho\right].
\end{equation}
Choosing the arbitrary parameter $\rho=L/\ell$, we obtain the well known
finite-size scaling result
\begin{equation}\label{universal}
X[t,g,\ell,L]=L^{\gamma_x/\nu}f\left(tL^{1/\nu}\right).
\end{equation}
Here the function $f(x)$ is a universal function of its argument. In
the remainder of this paper we will verify the scaling relation
(\ref{universal}) in the framework of model (\ref{model}).

\section{Finite-size scaling below the upper critical dimension}
\label{secfss}
\subsection{Method}
The method, we shall use here to analyze the finite-size scaling of the
model under consideration, is originally due to
L\"uscher~\cite{luscher82} in his study on the quantum ${\cal O}(n)$
nonlinear $\sigma$ model in $1+1$ dimensions. An extension of the method
was employed by Brezin and Zinn-Justin \cite{brezin85} and by Rudnick,
Guo, and Jasnow~\cite{rudnick85} in their works on the finite-size
scaling in systems with short-range potentials. Very recently it was
used in the investigation of crossovers in quantum ${\cal O}(n)$ systems
near their upper critical dimension~\cite{sachdev97}. We will see here
that the problem related to finite-size scaling in systems with
long-range forces can be successfully analyzed by the same approach.
Nevertheless, here we will observe the emergence of some subtleties,
which need to be discussed.

The central idea of the method is that at finite linear size $L$ of the
system, one can treat the ${\bbox k}=0$ mode of the field $\varphi(x)$,
playing the role of the magnetization, separately from the non zero
${\bbox k}$ modes. The non-zero modes are treated perturbatively
using the loop expansion. They are integrated out to yield an effective
Hamiltonian for the lowest mode only. All the modes being integrated out
are regulated in the infrared by $|{\bbox k}|^\sigma$ and consequently
the process is necessarily free of infrared divergences. On the other
hand the renormalizations of the bulk theory control the ultraviolet
divergences at finite size. In other words if we define by
\begin{equation}\label{magnetization}
\phi=\frac1V\int_V d^dx\varphi(x)
\end{equation}
the total spin by unit volume, then, from ${\cal H}$, we can get an
effective Hamiltonian function of $\phi$ after entirely integrating out
the $\varphi(k\neq0)$ fields:
\begin{equation}\label{effective}
{\cal H}_{\rm eff}=\frac{L^d}2\left(R\phi^2+\frac{U}2\phi^4\right).
\end{equation}

The coupling constants $R$ and $U$ are computed in powers of
$\varepsilon$, with the initial coupling constants renormalized as in
their bulk critical theory. This approach will rule out all the ultraviolet
divergences of the bulk critical point. The new coupling constants are
necessarily free of all ultraviolet divergences since the theory is
superrenormalizable~\cite{yamazaki77}. They are also free of infrared
divergences as we are only integrating out finite modes. Obviously,
these constants must obey the scaling forms,
\begin{equation}\label{scalingcoupling}
R=L^{\eta-2}f_R\left(tL^{1/\nu}\right) \ \ \ \ \text{and} \ \ \ \
U=L^{d-4+2\eta}f_U\left(tL^{1/\nu}\right)
\end{equation}
for $t\gtrsim0$, where $f_R$ and $f_U$ are scaling functions which
are properties of the bulk critical point. They are analytic at $t=0$.
This is a consequence of the fact that only finite modes have been
integrated out.

Once the scaling functions $f_R$ and $f_U$ are known one can attack the
problem of computing observables in the $\varphi^4$ theory with the
action ${\cal H}_{\rm eff}$. This theory is in dimension $d$ close to
the upper critical dimension $2\sigma$ (not in $d$ close to the usual
$4$), and the problem seems to be unsolvable. In the next section we
will show that it is not the case.

In order to investigate the long distance physics of the finite system,
one has to calculate thermal averages with respect to the new
effective Hamiltonian defined in~(\ref{effective}). They are
related to the thermodynamic functions of the system under
consideration. The averages of the field $\phi$ are defined by
\begin{equation}\label{averages}
{\cal M}_{2p}=\left<\left(\phi^2\right)^p \right>
=\frac{\int d^n\phi \ \phi^{2p}\exp\left(-{\cal H}_{\rm eff}\right)}
{\int d^n\phi \exp\left(-{\cal H}_{\rm eff}\right)},
\end{equation}
Using an appropriate rescaling of the field $\phi$:
$\Phi=\left(UL^d\right)^{1/4}\phi$, we can transform the effective
Hamiltonian into
\begin{equation}\label{neweff}
{\cal H}_{\rm eff}=\frac12z \Phi^2+\frac14\Phi^4,
\end{equation}
where the scaling variable $z=RL^{d/2}U^{-1/2}$ is an important
quantity in the investigations of finite-size scaling in critical
statics~\cite{brezin85,rudnick85} as well as in critical
dynamics~\cite{goldschmidt87,niel87}. With the effective Hamiltonian
(\ref{neweff}), we obtain the general scaling relation
\begin{equation}\label{scaling}
{\cal M}_{2p}=L^{-p(d-2+\eta)}\frac{L^{p(d-4+2\eta)/2}}{U^{p/2}}
f_{2p}\left(RL^{2-\eta} \frac{L^{(d-4+2\eta)/2}}{U^{1/2}}\right)
\end{equation}
for the momenta of the field $\phi$. Having in mind Eqs. 
(\ref{scalingcoupling}), we can write down Eq. (\ref{scaling}) in the 
following scaling form
\begin{equation}\label{Scaling}
{\cal M}_{2p}=L^{-p(d-2+\eta)}{\cal F}_{2p}(tL^{1/\nu}),
\end{equation}
in agreement with the finite-size scaling predictions of 
(\ref{universal}). In eq. (\ref{Scaling}), the function ${\cal
F}_{2p}(x)$ are universal.

%At the critical point $t=0$, we find
%that the scaling relation~(\ref{scaling}) reduces, up to the lowest 
%order in $\varepsilon$, to the very simple and useful form
%\begin{equation}\label{Scaling}
%{\cal M}_{2p}=\frac{L^{-p(d-2+\eta)}}{\left(U^*\right)^{p/2}}
%f_{2p}\left(\frac{R^*L^{2-\eta}}{\sqrt{U^*}}\right),
%\end{equation}
%where $R^*$ and $U^*$ means that the coupling constants $R$ and $U$
%are taken at the fixed point. Remark that instead of the variable
%$x=RL^{d/2}U^{-1/2}$, widely used in the literature
%\cite{brezin85,rudnick85,goldschmidt87,niel87}, one can use the
%argument of the scaling function $f_{2p}$ from equation~(\ref{Scaling}) 
%as a {\it characteristic variable} of the finite-size scaling properties
%of the model, i.e. one can define a new scaling variable $$
%z=RL^{2-\eta}U^{-1/2} $$ and evaluate it at the fixed point. Indeed this
%scaling variable is  $L$ independent at the critical point and will play
%a central role in our further analysis.

All the measurable thermodynamic quantities can be obtained from the
momenta ${\cal M}_{2p}$. For example the susceptibility is obtained
from
\begin{equation}\label{chi1}
\chi=\frac1n\int_V d^dx\left<\varphi(x)\varphi(0)\right>
=L^{2-\eta}{\cal F}_2(tL^{1/\nu}).
\end{equation}
Another quantity of importance for numerical analysis of the finite-size
scaling theory is the Binder's cumulant defined by
\begin{equation}\label{binder}
B=1-\frac13\frac{{\cal M}_4}{{\cal M}_2^2}.
\end{equation}

%At the bulk critical temperature $T_c$ the susceptibility and the
%Binder's cumulant are universal quantities.

In the remainder of this section we concentrate on the computation of
the coupling constants $R$ and $U$ of the effective Hamiltonian
(\ref{effective}) for the system with long-range interaction decaying
with the distance as a power law. As a consequence we will deduce
results for the characteristic variable
$z=RU^{-1/2}L^{2-\eta-\varepsilon/2}$, the susceptibility $\chi$ and the
amplitude ratio $r={\cal M}_4/{\cal M}_2^2$ entering the definition of
the Binder's cumulant.

\subsection{Computation of the coupling constants $R$ and $U$}
As we explained above, loop corrections will be treated perturbatively
on the non-zero ${\bbox k}$ modes. At the tree level (lowest order in
$\varepsilon=2\sigma-d$) this procedure generates a shift of the
critical temperature $T_c$ and a change of the coupling constant
$u_0$ and additional operators involving powers of $\varphi$ larger
than $4$. The calculations will be performed in the renormalized
theory. The renormalized coupling constant $u_R$ is expressed in
terms of the dimensionless coupling constant $g=\ell^\varepsilon u_R$
in which the parameter $\ell$ is an arbitrary length scale. Here we will
work in system in which $\ell=1$. Throughout these calculations we
use the minimal subtraction scheme. In this scheme, the counterterms
of the massless theory including the $\varphi^2$ insertions are
introduced. The one-loop counterterm for the coupling constant and
the $\varphi^2$ insertion will be the only one relevant in the lowest
corrections.

The finite-size correction to the renormalized coupling constant $t$
is given by
\begin{equation}\label{fssfort}
{\cal W}^t_{d,\sigma}(t,g,L)=(n+2)g \frac{1}{L^d}{\sum_{\bbox k}}'
\frac1{t+|{\bbox k}|^\sigma}
\end{equation}
to one-loop order.

In order to investigate the finite-size scaling of the model under
consideration one can use a suitable approach allowing to simplify the
analytical calculations. In the case $\sigma=2$ it is possible to
replace the summand by its Laplace transform. This is the so called
Schwinger representation. The aim of this approach is to reduce the
$d$-dimensional sum in the r.h.s of equation~(\ref{fssfort}) to the
one-dimensional effective problem. In the general case of arbitrary
$\sigma$, one cannot just use the Schwinger transformation or at least
in its familiar form. So we have to solve the problem by introducing
some kind of generalization for it. In the spirit of the same problem a
method to investigate the finite-size scaling in the framework of the
mean spherical model was suggested in Ref.~\cite{brankov89}. The method
is based upon the following genius identity
\begin{mathletters}\label{mittagide}
\begin{equation}
\frac1{1+z^\alpha}=\int_0^\infty dx \exp\left(-xz\right)x^{\alpha-1}
E_{\alpha,\alpha}\left(-x^\alpha\right),
\end{equation}
where the functions
\begin{equation}\label{schwinger}
E_{\alpha,\beta}(z)=\sum_{\ell=0}^\infty\frac{z^\ell}
{\Gamma\left(\alpha\ell+\beta\right)}
\end{equation}
\end{mathletters}
is the so called Mittag-Leffler type functions. For a more recent
review on these functions and other related to them, and their
application in statistical and continuum mechanics see
Ref.~\cite{mainardi97}. See  also Ref.~\cite{brankov89} and
Appendix~\ref{appA}.

Using the identity~(\ref{mittagide}), one gets, after some algebra,
\begin{mathletters}
\begin{equation}\label{jac}
{\cal W}^t_{d,\sigma}(t,g,L)=(n+2)g\frac{L^{\sigma-d}}{(2\pi)^\sigma}
\int_0^\infty dx x^{\frac\sigma2-1}E_{\frac\sigma2,\frac\sigma2}
\left(-x^{\sigma/2}\frac{tL^\sigma}{(2\pi)^\sigma}\right)
\left[{\cal A}^d(x)-1\right],
\end{equation}
where
\begin{equation}
{\cal A}(x)=\sum_{\ell=-\infty}^\infty e^{-x\ell^2}.
\end{equation}
\end{mathletters}
The analytic properties of the function ${\cal A}(x)$ are known very well.
For large $x$, ${\cal A}(x)-1$ decreases exponentially and the integral in
the r.h.s of Eq. (\ref{jac}) converges at infinity. For small $x$, the
Poisson transformation $ {\cal A}(x)=\left(\frac{\pi}{x}\right)^{\frac12}
{\cal A} \left(\frac{\pi^2}{x}\right)$ shows that ${\cal A}(x)$ converges.

For small $x$ the integral in the r.h.s of Eq. (\ref{jac}) has ultra
violet divergence for $\text{Re}\ d>\sigma$. So, an analytic continuation
in $d$ is required to give a meaning to the integral. Adding and
subtracting the small asymptotic behavior of the function ${\cal A}(x)$,
we get after some algebra
\begin{mathletters}\label{1step}
\begin{eqnarray}
{\cal
W}^t_{d,\sigma}(t,g,L)&=&(n+2)g\frac{L^{\sigma-d}}{(2\pi)^\sigma}
F_{d,\sigma}\left(tL^\sigma\right)\nonumber\\ && +2\pi(n+2)g
L^{\sigma-d}\left[(4\pi)^{d/2}\Gamma\left(\frac d2\right)\sigma
\sin\frac{d\pi}\sigma\right]^{-1}\left(tL^\sigma\right)^{d/\sigma-1},
\end{eqnarray}
where
\begin{equation}\label{It}
F_{d,\sigma}\left(y\right)=\int_0^\infty dx
x^{\frac\sigma2-1}E_{\frac\sigma2,\frac\sigma2}
\left(-\frac{yx^{\sigma/2}}{(2\pi)^\sigma}\right)
\left[{\cal A}^d(x)-1-\left(\frac{\pi}{x}\right)^{d/2}\right].
\end{equation}
\end{mathletters}
In the particular case $\sigma=2$, from Eq. (\ref{1step}) we recover
the result of Ref.~\cite{brezin85}.

By introducing the $\varphi^2$ counterterm insertion the renormalized
coupling constant $t$ is replaced by $tZ_t$, where $Z_t$ is given by
(\ref{Zt}). Hence to one loop order we have
\begin{equation}
R=t\left(1+\hat g\frac{n+2}\varepsilon\right)+
{\cal W}^t_{d,\sigma}(t,g,L).
\end{equation}

At $d=2\sigma$, ${\cal W}_{d,\sigma}^t(t,g,L)$ has a simple pole. An
expansion about this pole leads to the final expression
\begin{equation}\label{finalt}
R=t+\frac{n+2}\sigma\hat gt\ln t
+2^{\sigma-1}(n+2) \Gamma(\sigma)
\hat g L^{-\sigma}F_{2\sigma,\sigma}\left(tL^\sigma\right)+ {\cal
O}\left(\hat g^2\right).
\end{equation}
This result shows that, at the critical point, $R$ has the required
scaling properties of Eq.~(\ref{scalingcoupling}), since
\begin{equation}\label{nuex}
\nu^{-1}=\sigma-\frac{n+2}{n+8}\varepsilon+{\cal O}(\varepsilon^2).
\end{equation} 

For the finite system the renormalized coupling constant $g$, to
one-loop order, is shifted by a quantity expressed in the form
\begin{equation}\label{fssforg}
{\cal W}^g_{d,\sigma}(t,g,L)=-(n+8)g^2 \frac{1}{L^d}{\sum_{\bbox
k}}' \frac1{\left(t+|{\bbox k}|^\sigma\right)^2}.
\end{equation}
As one can see the summand here can be expressed as the first
derivative of  the summand of Eq.~(\ref{fssfort}) with respect to $t$. So,
the  result for $U$ can be derived from that of $R$. Using this fact one
gets
\begin{equation}
{\cal W}^g_{d,\sigma}(t,g,L)=(n+8)g^2
\left[\frac{L^{2\sigma-d}}{(2\pi)^{\sigma}}
F_{d,\sigma}'\left(tL^\sigma\right)-
L^{2\sigma-d}\frac2{\sigma(4\pi)^{d/2}}\frac{\Gamma\left(2-
d/\sigma\right)\Gamma\left(d/\sigma\right)}{\Gamma(d/2)}
\left(tL^\sigma\right)^{d/\sigma-2}\right],
\end{equation}
where the prime indicates that we have the derivative of the function $F$
with respect to its argument.

At the fixed point one ends up with
\begin{eqnarray}\label{finalg}
U&=& g\left[1+\hat g\frac{n+8}\sigma(1+\ln t)
+\hat g\frac{n+8}{2^{1-\sigma}}\Gamma(\sigma) F_{2\sigma,\sigma}'
\left(tL^\sigma\right)+ {\cal O}\left(\hat
g^2\right)\right]
\end{eqnarray}
for the renormalized coupling constant $U$. Eq. (\ref{finalg}) is
obtained using the fact that at one-loop order the coupling constant
is renormalized by $Z_g$ form Eq.~(\ref{Zg}). The obtained expression
(\ref{finalg}) shows that the coupling constant $U$ obeys the scaling
law of Eq. (\ref{scalingcoupling}). Note that $U$ has a finite limit as
$t\to0$, i.e. it is analytic at the bulk critical temperature. Indeed
as $t\to0$ one can use the expansion of the function
$F_{2\sigma,\sigma}(y)$ for small $y$ given by (see Appendix~\ref{appA})
\begin{mathletters}\label{expansion}
\begin{equation}\label{smally}
F_{2\sigma,\sigma}(y)=F_{2\sigma,\sigma}(0)+
2^{-\sigma}y {\cal C}_\sigma-\frac{2^{1-\sigma}}
{\sigma\Gamma(\sigma)}y\ln y+
{\cal O}(y^2),
\end{equation}
where
\begin{equation}
{\cal C}_\sigma=\frac1{\Gamma(\sigma)}\int_0^\infty\frac{du}u
\left[E_{\frac\sigma2,1}\left(-\frac{u^{\sigma/2}}{(2\pi)^\sigma}\right)
-\frac{u^\sigma}{\pi^\sigma}{\cal A}^{2\sigma}(u)+
\frac{u^\sigma}{\pi^\sigma}\right].
\end{equation}
\end{mathletters}
After substitution of~(\ref{expansion}) in~(\ref{finalg}) the terms
proportional to $\log y$ cancel, which shows that the coupling constant
$U$ is finite at $t=0$. Whence, one gets
$$
U=gL^{-\varepsilon}\left[1+\hat g\frac{n+8}\sigma\left(1+
\frac\sigma2\Gamma(\sigma){\cal C}_\sigma\right)+
{\cal O}(\hat g^2)\right] 
$$
showing that $U$ is analytic, as it should be, at the critical point.

\subsection{Some thermodynamic quantities}
\subsubsection{Shift of the critical point}
It is obvious that the coupling constant $R$ in the effective 
Hamiltonian (\ref{effective}) is just the deviation of the temperature
of the system from its `critical' value. By setting $t=0$ in
(\ref{finalt}), we obtain an expression for the finite-size shift of the
bulk critical temperature $T_c$. This is given by
\begin{equation}\label{fsshift}
T_c-T_c(L)= \varepsilon2^{\sigma-1}\frac{n+2}{n+8}\Gamma(\sigma)
L^{-\sigma}F_{2\sigma,\sigma}(0),
\end{equation}
where the coefficient $F_{2\sigma,\sigma}(0)$, appearing in the right
hand side of (\ref{fsshift}) can be evaluated for some particular values
of the interparticle interaction range $\sigma$
\cite{chamati001}
\begin{equation}\label{values}
F_{2\sigma,\sigma}\left(0\right)=\left\{
\begin{array}{ll}
2\zeta\left(1/2\right), &\sigma=1/2,\\[.5cm]
4\zeta\left(1/2\right)\beta\left(1/2\right), \ \ \ \ \ &\sigma=1,
\\[.5cm]
-4.82271993, &\sigma=3/2,\\[.5cm]
-8\ln2, &\sigma=2.
\end{array}
\right.
\end{equation}
Here $\zeta(x)$ is the Riemann zeta function with
$\zeta\left(\frac12\right)=-1.460354508 ...$ and $\beta(x)$ is the
analytic continuation of the Dirichlet series:
$$
\beta(x)=\sum_{\ell=0}^\infty\frac{(-1)^\ell}
{\left(2\ell+1\right)^{x}}, 
$$ 
with $\beta\left(\frac12\right)=0.667691457 ...$. Remark that the
function $F_{2\sigma,\sigma}(0)$ increases as the parameter
$\sigma$ vanishes.

In fact, since there is no true phase transition in the finite  system,
the critical temperature is shifted to a `pseudocritical'  temperature, 
$T_c(L)$, corresponding to the rounding of the thermodynamic
singularities holding in the bulk limit. From (\ref{fsshift}) one remarks
that $T_c(L)$ is larger than $T_c$, confirming previously obtained
results in the framework of the spherical model~\cite{chamati96}. 
Notice also that for the shift exponent $\lambda$, we get 
$\lambda=\sigma$ to lowest order in $\varepsilon$.

\subsubsection{Binder's cumulant}
In this subsection we are interested in the calculation of the
amplitude ratio $r={\cal M}_4/{\cal M}^2_2$ instead of the Binder's
cumulant from definition (\ref{binder}). This quantity can be expressed
in power series of the scaling variable $z=RL^{2-\eta-\varepsilon/2}U^{-1/2}$ as
\begin{eqnarray}\label{amplitude}
r&=&\frac n4\frac{\Gamma^2\left(\frac14n\right)}
{\Gamma^2\left(\frac14(n+2)\right)}\left\{1-
z\left[\frac{\Gamma\left(\frac14(n+6)\right)}
{\Gamma\left(\frac14(n+4)\right)}+
\frac{\Gamma\left(\frac14(n+2)\right)}
{\Gamma\left(\frac14n\right)}
-2\frac{\Gamma\left(\frac14(n+4)\right)}
{\Gamma\left(\frac14(n+2)\right)}\right]\right.\nonumber\\
&&\left.+z^2\left[\frac{\Gamma\left(\frac14(n+6)\right)
\Gamma\left(\frac14(n+2)\right)}
{\Gamma\left(\frac14(n+4)\right)\Gamma\left(\frac14n\right)}
+3\frac{\Gamma^2\left(\frac14(n+4)\right)}
{\Gamma^2\left(\frac14(n+2)\right)}-n-1\right] +{\cal
O}\left(z^3\right)\right\}.
\end{eqnarray}
So, in order to obtain a result for $r$ it is enough to evaluate $z$ at
the fixed point $g^*$ and to deduce the value for the Binder's cumulant.
As we mentioned before this parameter appear in all thermodynamic
functions through the momenta defined earlier in this paper.

At the fixed point $g^*$ in the vicinity of the upper critical
dimension, we obtain
\begin{eqnarray}\label{zfixed}
z^*&\equiv& \left.\frac{R L^{2-\eta}}{\sqrt{UL^\varepsilon}}
\right|_{\rm fixed point}\nonumber\\
&=&\frac1{\sqrt{g^*}}\left[y-\frac\varepsilon{2\sigma}y
\left(1-\frac{n-4}{n+8}\ln y\right)+2^{\sigma-1}\varepsilon
\frac{n+2}{n+8}\Gamma(\sigma)F_{2\sigma,\sigma}\left(y\right)
\right.\nonumber\\
& &\left.-\varepsilon2^{\sigma-2}y\Gamma(\sigma)
F_{2\sigma,\sigma}'\left(y\right)\right].
\end{eqnarray}
This result is obtained by using~(\ref{nuex}) and the fact that up to
one loop order the terms proportional to $\ln L$ cancel. In
Eq.~(\ref{zfixed}), we introduce the scaling variable $y=tL^{1/\nu}$.
Finally let us notice that from this equation one can see easily that
$z^*$ verifies the finite-size scaling hypotheses and consequently all
the thermodynamic functions do.

At the critical temperature $T_c$ (i.e. $t=0$, and so $y=0$), we obtain
\begin{equation}\label{z0}
z_0^*=\sqrt\varepsilon\left[\frac{n+2}{\sqrt{n+8}}\sqrt{
\frac{\Gamma(\sigma)}{2\pi^\sigma}}F_{2\sigma,\sigma}(0)+
{\cal O}(\varepsilon)\right].
\end{equation}

Numerical values for the amplitude ratio (\ref{amplitude}) can be
obtained by replacing the value of $z_0^*$ form (\ref{z0}) and taking
some specific values of the small parameter $\varepsilon$. Note that the
scaling variable $z$ is proportional to $\sqrt\varepsilon$ as it was
found previously (see Ref. \cite{brezin85} for example) in the case of
short-range forces. Furthermore it coincides with the result of
Ref.~\cite{luijten99} for the scaling variable $x$ in the case of
long-range interaction. Consequently all the thermodynamic function will
be computed in powers of $\sqrt\varepsilon$.

\subsubsection{Magnetic Susceptibility}
As we mentioned earlier, there is no phase transition in the finite 
system under consideration. Consequently there will be no `true' 
correlation length. An expression for it can be deduced from that of 
the susceptibility (\ref{chi1}) trough the relation:
\begin{equation}
\xi^{2-\eta}=\chi.
\end{equation}
The analyticity of the susceptibility is a consequence of that the 
coupling constants $R$ and $U$.

From (\ref{chi1}) in the region $tL^\sigma\ll1$ (i.e. $z\ll1$), we obtain
for the susceptibility
\begin{eqnarray}
\chi&=&\frac{L^\sigma}{\sqrt\varepsilon}
\frac{2\sqrt2}{\sqrt{(4\pi)^\sigma\Gamma(\sigma)}}
\frac{\sqrt{n+8}}{n}\frac{\Gamma\left(\frac14(n+2)\right)}
{\Gamma\left(\frac n4\right)}\left[1-z\left(
\frac n4\frac{\Gamma\left(\frac n4\right)}
{\Gamma\left(\frac14(n+2)\right)}
-\frac{\Gamma\left(\frac14(n+2)\right)}
{\Gamma\left(\frac n4\right)}\right)\right.\nonumber\\
&&\left.+z^2\left(\frac{1-n}4+\frac{\Gamma^2\left(
\frac14(n+2)\right)}{\Gamma^2\left(\frac n4\right)}\right)
-\hat g\frac{n+8}\sigma\left(1+\frac\sigma2\Gamma(\sigma)
{\cal C}_\sigma\right)+{\cal O}(\hat g z,z^3)\right]
\end{eqnarray}
at the bulk critical point $T_c$. 

To the lowest order in $\varepsilon$, after taking the limit $n\to\infty$,
we find that the correlation length scales like 
$$
\xi\sim \varepsilon^{-1/2\sigma} L,
$$
confirming the results obtained in the spherical model~\cite{brankov90} 
and showing that this behavior is not a characteristic of the 
spherical limit i.e. $n\to\infty$.

In the region $tL^\sigma\gg1$ (i.e. $z\gg1$), 
%to obtain the finite-size 
%corrections to the bulk susceptibility $\chi_\infty$ one needs the 
%identity
%$$
%\int_0^\infty x^{2\beta-1} x^{2\beta-1} e^{-zx^2-x^4}=
%\frac{\Gamma(\beta)}{2z^\beta}\left[1+\beta(\beta+1)z^{-2}
%+{\cal O}\left(z^{-4}\right)\right], \ \ \ \beta>0.
%$$ 
%With the help of this identity 
from Eq.~(\ref{chi1}), we get
\begin{eqnarray}\label{correction}
\chi&=&\frac1t\left[1-\frac{n+2}\sigma\hat g\ln t-2^{\sigma-1}(n+2) 
\Gamma(\sigma)\hat g \left(tL^\sigma\right)^{-1}
F_{2\sigma,\sigma}\left(tL^\sigma\right)\right.\nonumber\\
&&\left.-\frac12(n+2)(4\pi)^\sigma\Gamma(\sigma)\hat g
\left(tL^\sigma\right)^{-2}+{\cal O}\left(\hat g^2\right)\right].
\end{eqnarray}
The function $F_{d,\sigma}(y)$ has the following large $y$ asymptotic  
behavior (see Appendix \ref{appB})
\begin{mathletters}\label{largey}
\begin{equation}\label{largeya}
F_{d,\sigma}(y)\simeq-\frac{(2\pi)^\sigma}y+
\frac{4^\sigma\pi^{\sigma-d/2}\Gamma\left(\frac{d+\sigma}2\right)}
{y^2\Gamma(-\frac\sigma2)}{\sum_{\bbox l}}'\frac1{|\bbox l|^{d+\sigma}}
\end{equation}
for the case $0<\sigma<2$, and
\begin{equation}\label{largeyb}
F_{d,2}(y)\simeq-\frac{4\pi^2}y+d(2\pi)^{(5-d)/2}y^{(d-3)/4}e^{-\sqrt y}
\end{equation}
for the particular case $\sigma=2$.
\end{mathletters}
These results show that the last term in Eq. (\ref{correction}) is 
just canceled by the first term in Eqs. (\ref{largey}).

In the case of long-range interaction $0<\sigma<2$, we obtain for the 
susceptibility
\begin{equation}\label{longrange}
\chi=\chi_\infty\left[1-\sigma\hat g(n+2)2^{3\sigma-2} 
\left(tL^\sigma\right)^{-3}
\frac{\Gamma\left(3\sigma/2\right)\Gamma(\sigma)}
{\Gamma(1-\sigma/2)}{\sum_{\bbox l}}' \
{\bbox l}^{-3\sigma}+{\cal O}(\hat g^2)\right]
\end{equation}
in agreement with the finite-size scaling hypothesis (\ref{chi1}). 
Eq. (\ref{longrange}) shows that the finite-size scaling behavior of 
the system is dominated by the bulk critical behavior, with small 
correction in powers of $L$. It should be noted that the above result 
cannot be continued smoothly to the case of short-range interaction 
$\sigma=2$, since then $F_{4,2}(y)$ (see Eq. (\ref{largeyb})) falls 
off exponentially fast and, correspondingly, the the finite-size 
corrections to $\chi$ are exponentially small:
\begin{equation}
\chi=\chi_\infty\left[1-8\hat g\sqrt{2\pi}(n+2)\left(tL^2\right)^{-3/4}
e^{-\sqrt t L}\right]
\end{equation}
At this point we are in disagreement with the statement given in 
Ref.~\cite{chen2000} that the approach used in Refs. 
\cite{luijten99,brezin85,rudnick85,sachdev97,goldschmidt87,niel87}
yields an incorrect non-exponential result. Note that all our
calculations are up to the order $\hat g^1$. It is interesting to see what
happen in higher order, e.g. $\hat g^2$, in this case, however, we need
to have at our disposal the corresponding high order terms in Eqs.
(\ref{finalt}) and (\ref{finalg}). Indeed it is beyond the scope of the
present study. First the power law fall off of the finite-size corrections
to the bulk critical behavior, due to long-range nature of the 
interaction, was found in the framework of the spherical 
model~\cite{singh89,brankov91}. Here, we extended this result to finite
$n$ using a perturbative approach. 

%In the region $tL^\sigma\gg1$ (i.e. $z\gg1$) corresponding to large
%system sizes compared to the reduced temperature, from
%Eq.~(\ref{finalt}), we have
%\begin{equation}\label{correction}
%\xi^{-\sigma}=t+\varepsilon\sigma2^{3\sigma-2} t^{-2}L^{-3\sigma}
%\Gamma\left(3\sigma/2\right)
%\frac{\Gamma(\sigma)}{\Gamma(1-\sigma/2)}{\sum_{\bbox l}}' \
%{\bbox l}^{-3\sigma}.
%\end{equation}
%This result shows that the finite-size scaling behavior of the system is
%dominated by the bulk critical behavior, with small corrections in
%powers of $L$. It should be noted that it cannot be continued smoothly
%to the case $\sigma=2$, since then $E_{\sigma/2,\sigma/2}(-x)$ falls off
%exponentially fast and, correspondingly, the finite-size corrections to
%$\xi^{-\sigma}$ are expected to be exponentially small. Such a
%behaviour of the power law fall-off of the finite-size corrections was
%found for the special case of the spherical model model~\cite{singh89}.
%Expression (\ref{correction}) is an extension, for finite $n$, using
%$\varepsilon$ expansion. Note that  if we set $\sigma=2$ 
%in~(\ref{finalt}), we would find an exponential decay of the 
%finite-size corrections to the susceptibily in agreement with Ref. 
%\cite{sachdev97}. This result is in disagreement with Ref. 
%\cite{chen2000}, where a power law for the corrections was found.

\section{conclusions}\label{discussion}
In this paper, we have investigated the finite-size scaling properties
in the ${\cal O}(n)$-symmetric $\varphi^4$ model with long-range
interaction potential decaying algebraically with the interparticle
distance. We have found that the methods developed in
Refs.~\cite{luscher82,brezin85,rudnick85,sachdev97} can be successfully
extended to systems with long-range interaction by combining them with
other known techniques. These techniques allow the investigation to be
simplified and express the results for various thermodynamic functions
in terms of simple and known mathematical functions.

Here we restricted our calculations to the critical domain $T\gtrsim
T_c$ and investigated the model in dimensions less than the upper
critical one, which turns out to be $2\sigma$ ($0<\sigma\leq2$). We
constructed an  effective Hamiltonian, from the initial one, with new
coupling  constants $R$ and $U$. These constants obey the scaling
hypothesis~(\ref{scalingcoupling}). We found that the even momenta of
the field $\varphi$, related to the thermodynamics of the finite system,
are scaling functions of the characteristic variable $$
z=RU^{-1/2}L^{2-\eta-\varepsilon/2} . $$ This variable has the required
scaling form predicted by the finite-size scaling theory. From the
obtained forms of the constant $R$ and $U$ one concludes that $z$ is a
universal quantity, which does not depend of the details of the model.

We evaluated the finite-size shift, the susceptibility and the
amplitude ratio $r={\cal M}_4/{\cal M}_2^2$ at the tree level (lowest
order in $\varepsilon  $). We observed that the critical behavior of the
system is dominated by its bulk critical behavior away from the critical
domain and that the finite-size scaling is relevant in the vicinity of
the critical point. The amplitude ratio $r$ is evaluated as an expansion
in powers of $z\sim\sqrt\varepsilon$. Our result is in consistency with
that of Ref.~\cite{luijten99}. But it is disagreement with the numerical
results of the same paper. There, it has been found, using the Monte
Carlo method, that $r$ has an expansion in $\varepsilon$ instead of its
square root. At this time, we do not have a reasonable explanation of
this fact. It is also possible that higher order in $\varepsilon$ could
improve the result. An amelioration of the  result could also come from
accounting finite cutoff effects, which were to  be relevant in the
investigation of finite systems and the comparison  of the results with
numerical works \cite{chen991,chen992}. However this is the subject of 
another publication.

Notice that in the only work devoted to the exploration of finite-size
scaling in ${\cal O}(n)$ systems with long-range interaction
(Ref.~\cite{luijten99}) the pertinent integrals have to be evaluated
only numerically, due to the choice of a parametrization that does not
reduce the  $d$-dimesional problem to the effective one dimensional one.
The approach we used here is more efficient in the sense that the 
corresponding final expressions can be handled by analytical means.
Consequently, we cannot make a direct comparison between the results of 
this paper and those obtained there.

%was limited strictly to the critical point i.e. $t=0$. In this case it is
%easy to take care about the mathematical complications arising from
%the presence of the parameter $\sigma$ controlling the range of the
%interaction. The results we obtained here are more general in the
%sense that they are valid also for $t\neq0$.

Let us note that it would be interesting and useful to extend the result
obtained here in the static limit to models including dynamics, since
we believe that this is closely related to the extensively investigated
filed of quantum critical points i.e. phase transitions occurring at
zero-temperature. In particular we find it useful to investigate the
critical dynamic of the quantum model considered in
Ref~\cite{chamati00} in the large $n$ limit.

\acknowledgments
The authors thank Dr. E. Koroutcheva for helpful discussion. This work
is supported by The Bulgarian Science Foundation under Project F608.

\appendix
\section{Some properties of the Mittag-leffler type functions}
\label{appA}
The Mittag-leffler type functions are defined by
the power series~\cite{mainardi97}:
\begin{equation}\label{mittag}
E_{\alpha,\beta}(z)=\sum_{k=0}^\infty\frac{z^k}{\Gamma(\alpha k+\beta)},
\ \ \ \alpha,\beta>0.
\end{equation}
They are entire functions of finite order of growth. Let us mention
that the function corresponding the particular case $\beta=1$ was
introduced by Mittag-Leffler. These function are very popular in the
field of fractional calculus (for a recent review see
Ref.~\cite{mainardi97}).

One of the most striking properties of these functions is that they
obey the following useful identity~\cite{mainardi97})
\begin{equation}\label{ide}
\frac1{1+z}=\int_0^\infty dx e^{-x}x^{\beta-1}E_{\alpha,\beta}
\left(-x^\alpha z\right),
\end{equation}
which is obtained by means of term-by-term integration of the
series~(\ref{mittag}). The integral in Eq.~(\ref{ide}) converges in the
complex plane to the left of the line $\text{Re} z=1^{1/\alpha}$, $|\arg
z|\leq\frac12\alpha\pi$. The identity (\ref{ide}) lies in the basis of the
mathematical investigation of finite-size scaling in the spherical model
with algebraically decaying long-range interaction (see Ref.
\cite{brankov92} and references therein).

In some particular cases the functions $E_{\alpha,\beta}(z)$ reduces
to known functions. For example, in the case corresponding to the
short range case we have:
\begin{equation}\label{simple}
E_{1,1}(z)=\exp(z).
\end{equation}

Setting $z=y^{-\alpha}$, $y>0$, and $x=ty$, we obtain the Laplace
transform
\begin{equation}\label{ide1}
\frac{y^{\alpha-\beta}}{1+z^\alpha}=\int_0^\infty dt e^{-zt}t^{\beta-1}
E_{\alpha,\beta}\left(-t^\alpha\right)
\end{equation}
from which we derive the identity (\ref{mittagide}) by setting
$\beta=\alpha$.

The asymptotic behavior of the Mittag-Leffler functions is given by
the Lemma~\cite{bateman55}:

Let $0<\alpha<2$, $\beta$ be an arbitrary complex number, and $\gamma$
be a real number obeying the condition
$$
\frac12\alpha\pi<\gamma<\min\{\pi,\alpha\pi\}.
$$
Then for any integer $p\geq1$ the following asymptotic expressions
hold when $|z|\to\infty$:
\begin{itemize}
\item At $|\arg z|\leq\gamma$,
\begin{equation}
E_{\alpha,\beta}(z)=\frac1\alpha z^{(1-\beta)/\alpha}e^{z^{1/\alpha}}
-\sum_{k=1}^\infty\frac{z^{-k}}{\Gamma(\beta-\alpha k)} +{\cal
O}\left(|z|^{-p-1}\right).
\end{equation}
\item At $\gamma\leq|\arg z|\leq\pi$,
\begin{equation}\label{largez}
E_{\alpha,\beta}(z)=
-\sum_{k=1}^\infty\frac{z^{-k}}{\Gamma(\beta-\alpha k)} +{\cal
O}\left(|z|^{-p-1}\right).
\end{equation}
\end{itemize}

\section{Asymptotic behavior of the function 
$F_{\lowercase{d},\sigma}(\lowercase{y})$}\label{appB}
To obtain the small $y$ behavior~(\ref{smally}) of the function
$F_{d,\sigma}(y)$ we use the identity~\cite{brankov89}
\begin{equation}
\ln\phi=\alpha\int_0^\infty\frac{dx}x\left[
E_{\alpha,1}\left(-x^\alpha\right)-
E_{\alpha,1}\left(-yx^\alpha\right)\right]
\end{equation}
and the definition of the function $F_{d,\sigma}(y)$:
\begin{equation}\label{b2}
F_{d,\sigma}\left(y\right)=\int_0^\infty dx
x^{\frac\sigma2-1}E_{\frac\sigma2,\frac\sigma2}
\left(-\frac{yx^{\sigma/2}}{(2\pi)^\sigma}\right)
\left[{\cal A}^d(x)-1-\left(\frac{\pi}{x}\right)^{d/2}\right].
\end{equation}
After some algebra one obtains:
\begin{mathletters}
\begin{equation}
F_{2\sigma,\sigma}(y)=F_{2\sigma,\sigma}(0)+
2^{-\sigma}y {\cal C}_\sigma-\frac{2^{1-\sigma}}
{\sigma\Gamma(\sigma)}y\ln y+
{\cal O}(y^2),
\end{equation}
where
\begin{equation}
{\cal C}_\sigma=\frac1{\Gamma(\sigma)}\int_0^\infty\frac{du}u
\left[E_{\frac\sigma2,1}\left(-\frac{u^{\sigma/2}}{(2\pi)^\sigma}\right)
-\frac{u^\sigma}{\pi^\sigma}{\cal A}^{2\sigma}(u)+
\frac{u^\sigma}{\pi^\sigma}\right].
\end{equation}
\end{mathletters}

To obtain the large $y$ asymptotic behavior~(\ref{largey}) of the function 
$F_{d,\sigma}(y)$ we rewrite (\ref{b2}) in the form
\begin{equation}\label{other}
F_{d,\sigma}(y)=\pi^{d/2}\int_0^\infty dx x^{\frac\sigma2-\frac d2-1}
E_{\frac\sigma2,\frac\sigma2}\left(-\frac{yx^{\sigma/2}}{(2\pi)^\sigma
}\right){\sum_{\bbox l}}'e^{-\pi^2{\bbox l}^2/x}-\int_0^\infty
dx x^{\frac\sigma2-1}
E_{\frac\sigma2,\frac\sigma2}\left(-\frac{yx^{\sigma/2}}{(2\pi)^\sigma
}\right).
\end{equation}
Using the identity 
\begin{equation}
\int_0^\infty
dx x^{\frac\sigma2-1}
E_{\frac\sigma2,\frac\sigma2}\left(-x^{\sigma/2}\right)=1, \ \ \ \ 
\sigma>0
\end{equation}
From the second term of Eq.~(\ref{other}) we obtain the first terms of 
Eqs.~(\ref{largeya}) and (\ref{largeyb}) respectively.

Next taking into account Eq. (\ref{largez}) or Eq. (\ref{simple}) for 
the function $E_{\alpha,\beta}(z)$ and after subsequent integration 
in the first term of Eq. (\ref{other}), we obtain finally the asymptotic 
behavior given by Eqs. (\ref{largeya}) and (\ref{largeyb}).

\end{document}